\documentclass[10pt]{article} %10pt, y para trabajar 12pt
\usepackage{amssymb}
\begin{document}
\setlength{\baselineskip}{13pt} %13pt, y para trabajar 16pt
\begin{center}
{\Large {\bf \mbox{} \\ \mbox{} \\ An integrable family of 
Poisson systems: \vspace{2mm} \\ characterization and global analysis}}
\end{center}

\mbox{}

\begin{center}
{\large Benito Hern\'{a}ndez--Bermejo \footnote{\normalsize {\em E-mail address:\/} 
{\tt benito.hernandez@urjc.es } \\ \mbox{} \hspace{0.35cm} Telephone: (+34) 91 488 73 91. 
Fax: (+34) 91 488 73 38.}  } \\
\mbox{} \\
Departamento de Matem\'{a}tica Aplicada. Universidad Rey Juan Carlos. \\
E.S.C.E.T. Campus de M\'{o}stoles. Edificio Departamental II. \\
Calle Tulip\'{a}n S/N. 28933--M\'{o}stoles--Madrid. Spain. \\
\end{center}

\mbox{}

\mbox{}

\noindent \rule{15.6cm}{0.01in}
\noindent{\bf Abstract} \vspace{2mm}

A family of solutions of the Jacobi PDEs is investigated. This family is $n$-dimensional, of arbitrary nonlinearity and can be globally analyzed (thus improving the usual local scope of Darboux theorem). As an outcome of this analysis it is demonstrated that such Poisson structures lead to integrable systems. The solution family embraces as particular cases different systems of applied interest that are also regarded as examples.

\mbox{}


\noindent \rule{15.6cm}{0.01in}

\mbox{}

\mbox{}

The search and analysis of solutions of the Jacobi partial differential equations [\ref{olv1}],
\begin{equation}
     \label{z1jac}
     \sum_{l=1}^n ( J_{li} \partial_l J_{jk} + J_{lj} \partial_l J_{ki} + 
     J_{lk} \partial_l J_{ij} ) = 0 \:\; , \;\:\;\: i,j,k=1, \ldots ,n
\end{equation}
such that the so-called structure functions $J_{ij}(x_1, \ldots ,x_n) \equiv J_{ij}(\mathbf{x})$ must also verify an additional skew-symmetry condition, 
\begin{equation}
     \label{z1sksym}
     J_{ij} =  - J_{ji} \:\; , \;\:\;\: i,j=1, \ldots ,n
\end{equation}
has deserved an important attention during the last decades (for instance, see [\ref{bsn1}] and references therein). In (\ref{z1jac}-\ref{z1sksym}), the $C^{\infty}$ functions $J_{ij}(\mathbf{x})$ conform the entries of an $n \times n$ structure matrix ${\cal J}(\mathbf{x})$ which can be degenerate in rank. The interest of this problem arises from its application in the framework of Poisson systems. Recall that a finite-dimensional dynamical system is said to have a Poisson structure if it can be written in the form 
\begin{equation}
\label{z1nham}
     \frac{\mbox{\rm d} x_i}{\mbox{\rm d} t} = \sum_{j=1}^n J_{ij}(\mathbf{x}) 
	\partial _j H(\mathbf{x}) \; , \;\:\; i = 1, \ldots , n 
\end{equation} 
or briefly $\dot{\mathbf{x}}= {\cal J}(\mathbf{x}) \cdot \nabla H(\mathbf{x})$, where $ \partial_j \equiv \partial / \partial x_j$, function $H$ is a time-independent first integral and ${\cal J}$ is a structure matrix. Finite-dimensional Poisson systems (see [\ref{olv1}] and references therein for an overview and a historical discussion) are ubiquitous in most fields of applied mathematics. Moreover, describing a given system in terms of a Poisson structure allows the obtainment of a wide range of information thanks to a plethora of specially adapted methods developed for such kind of systems. For instance, in [\ref{bsn1}] some references about applications in diverse domains and tools associated to the Poisson formulation can be found. Apart from providing a wide formal generalization of classical Hamiltonian systems (for instance, allowing for odd-dimensional vector fields) Poisson systems have interest after their (at least local) dynamical equivalence with classical Hamiltonian equations, as stated by Darboux theorem [\ref{olv1}]. This justifies that Poisson systems can be regarded, to a large extent, as a rightful generalization of classical Hamiltonian systems. For these reasons, the investigation of the set of nonlinear PDEs (\ref{z1jac}-\ref{z1sksym}) is of central interest in this context [\ref{gyn1}], in particular the search of solutions that are defined for arbitrary dimension $n$ in terms of functions of arbitrary nonlinearity [\ref{byq}] and such that the Darboux reduction can be carried out globally [\ref{bsn1},\ref{byv4}] (this is, improving the {\em a priori \/} scope of Darboux theorem). The last condition is due to the fact that the global determination of the Darboux coordinates is usually a nontrivial task only known for a limited sample of Poisson structures. In this work, a new result of this kind is developed. We begin with a first statement:

\mbox{}

\noindent{\bf Theorem 1.} 
{\em Let  $\eta (\mathbf{x})$ and $\varphi_i(x_i)$, for $i=1, \ldots ,n$, be functions defined in a domain $\Omega \subset \mathbb{R}^n$, all of which are $C^{\infty}(\Omega)$ and nonvanishing in $\Omega$. Let $\kappa _{ij}$, $i,j = 1, \ldots , n$, be arbitrary real constans that are skew-symmetric 
\[
	\kappa _{ij} + \kappa _{ji}=0 \: , \:\:\: i,j=1, \ldots ,n
\]
and satisfy the zero-sum conditions
\begin{equation}
\label{z1kappa2}
    \kappa _{ij} + \kappa _{jk} + \kappa _{ki} = 0 \: , \:\:\: i,j,k=1, \ldots ,n
\end{equation}
In addition let 
\begin{equation}
\label{z1psi}
	\psi_i(x_i) = \int \frac{\mbox{d}x_i}{\varphi_i(x_i)}\: , \:\:\: i=1, \ldots ,n
\end{equation}
denote one primitive of $1/ \varphi_i(x_i)$. Finally, let the functions 
$\chi_{ij}(x_i,x_j)$ be defined by
\[
	\chi_{ij}(x_i,x_j) = \psi_i(x_i) - \psi_j(x_j) + \kappa_{ij} 
	\: , \:\:\: i,j=1, \ldots ,n
\]
and assume that $\chi_{ij}(x_i,x_j)$ is nonvanishing in $\Omega$ at least for one pair $(i,j)$. Then ${\cal J}=(J_{ij})$ is a family of $n$-dimensional Poisson structure matrices globally defined in $\Omega$, where }
\begin{equation}
\label{z1sol1}
	J_{ij}(\mathbf{x})= \eta (\mathbf{x}) \varphi_i(x_i) \varphi_j(x_j) \chi_{ij}(x_i,x_j)
	\:\: , \:\:\:\:\: i,j = 1, \ldots ,n
\end{equation}

\mbox{}

\noindent{\bf Proof.} Skew-symmetry is evident in (\ref{z1sol1}). We then substitute ${\cal J}$ in (\ref{z1sol1}) into the Jacobi identities (\ref{z1jac}) and obtain after some algebra:
\[
     \sum_{l=1}^n ( J_{li} \partial_l J_{jk} + J_{lj} \partial_l J_{ki} + 
     J_{lk} \partial_l J_{ij} ) = \eta T_1 + \eta^2 T_2
\]
where $T_1$ and $T_2$ are the following terms, to be examined separately:
\[
	T_1 = \sum_{l=1}^n \varphi_i \varphi_j \varphi_k \varphi_l (\partial_l \eta)
	( \chi_{il} \chi_{jk} + \chi_{jl} \chi_{ki} + \chi_{kl} \chi_{ij} )
\]
\[
	T_2 = \sum_{l=1}^n \left\{ \varphi_i \varphi_l \chi_{il} \left[ \delta_{lj} 
	\varphi_j^{\prime} \varphi_k \chi_{jk} + \delta_{lk} \varphi_j \varphi_k^{\prime} 
	\chi_{jk}+\varphi_j \varphi_k \left( \delta_{lj}\frac{1}{\varphi_j}- 
	\delta_{lk}\frac{1}{\varphi_k} \right) \right] \right. +
\]
\[
	\varphi_j \varphi_l \chi_{jl} \left[ \delta_{lk} 
	\varphi_k^{\prime} \varphi_i \chi_{ki} + \delta_{li} \varphi_k \varphi_i^{\prime}
	\chi_{ki}+\varphi_k \varphi_i \left( \delta_{lk}\frac{1}{\varphi_k}- 
	\delta_{li}\frac{1}{\varphi_i} \right) \right] +
\]
\[
	\left. \varphi_k \varphi_l \chi_{kl} \left[ \delta_{li} 
	\varphi_i^{\prime} \varphi_j \chi_{ij} + \delta_{lj} \varphi_i \varphi_j^{\prime}
	\chi_{ij}+\varphi_i \varphi_j \left( \delta_{li}\frac{1}{\varphi_i}- 
	\delta_{lj}\frac{1}{\varphi_j} \right) \right] \right\}
\]
Regarding $T_1$, if every $\chi_{ij}$ is substituted by its expression $\chi_{ij}=\psi_i-\psi_j+\kappa_{ij}$ and the result is simplified, it is found that:
\[
	\chi_{il} \chi_{jk} + \chi_{jl} \chi_{ki} + \chi_{kl} \chi_{ij} = 
\]
\[
	(\kappa_{jk}+\kappa_{kl}-\kappa_{jl}) \psi_i + 
	(\kappa_{ki}+\kappa_{il}-\kappa_{kl}) \psi_j + 
	(\kappa_{ij}+\kappa_{jl}-\kappa_{il}) \psi_k -
\]
\[
	(\kappa_{jk}+\kappa_{ki}+\kappa_{ij}) \psi_l + 
	\kappa_{il} \kappa_{jk} + \kappa_{jl} \kappa_{ki} + \kappa_{kl} \kappa_{ij}
\]
In the last identity, the terms multiplied by one of the $\psi_1, \ldots , \psi_n$ vanish due to the zero-sum relations (\ref{z1kappa2}). In addition, using the same equations (\ref{z1kappa2}) we have:
\[
	\kappa_{il} \kappa_{jk} + \kappa_{jl} \kappa_{ki} + \kappa_{kl} \kappa_{ij} =
	\kappa_{il}(\kappa_{jl}-\kappa_{kl}) + \kappa_{jl}(\kappa_{kl}-\kappa_{il}) + 
	\kappa_{kl}(\kappa_{il}-\kappa_{jl})=0
\]
It is thus demonstrated that $T_1=0$. We proceed now with $T_2$: expanding its expression and cancelling out similar terms, after a suitable rearrangement we arrive at: 
\[
	T_2 = 2 \varphi_i \varphi_j \varphi_k [\chi_{ij}+\chi_{jk}+\chi_{ki}] = 
	2 \varphi_i \varphi_j \varphi_k [\kappa_{ij}+\kappa_{jk}+\kappa_{ki}] = 0
\]
Therefore it is also $T_2=0$ and the proof is complete. \hfill Q.E.D.

\mbox{}

Recall that, as indicated in the theorem, for every $i$ the primitive $\psi_i(x_i)$ obtained from $\varphi_i(x_i)$ in (\ref{z1psi}) must be chosen to be one and the same for all the entries of ${\cal J}$. However, the specific choice is actually arbitrary. It is also worth observing that the definition (\ref{z1psi}) allows an alternative expression for the family just characterized, namely ${\cal J}=(J_{ij})$ can also be written as
\[
	J_{ij}(\mathbf{x})= \frac{\eta (\mathbf{x})}{\psi_i^{\prime}(x_i)\psi_j^{\prime}(x_j)}\chi_{ij}(x_i,x_j) 
	= \frac{\eta (\mathbf{x})}{\psi_i^{\prime}(x_i)\psi_j^{\prime}(x_j)}
	[\psi_i(x_i) - \psi_j(x_j) + \kappa_{ij}] \:\: , \:\:\:\:\: i,j = 1, \ldots ,n
\]
where functions $\psi_i^{\prime}(x_i)$ are $C^{\infty}(\Omega)$ and nonvanishing in $\Omega$, while the rest of defining properties were already presented in Theorem 1. This can be taken as an alternative definition of the solution family of structure matrices. We can now characterize some of the properties of the family identified in Theorem 1: 

\mbox{}

\noindent {\bf Theorem 2.} 
{\em Let ${\cal J}$ be a structure matrix of the form (\ref{z1sol1}) characterized in Theorem 1, which is defined in a domain $\Omega \subset \mathbb{R}^n$ and such that the pair $(i,j)$ verifies that function $\chi_{ij}(x_i,x_j)$ is nonvanishing in $\Omega$. 
Then Rank(${\cal J}$)$=2$ everywhere in $\Omega$ and a complete set of functionally independent Casimir invariants for ${\cal J}$ is given by:}
\begin{equation}
\label{z1cas}
	C_{k}(\mathbf{x}) = \frac{\psi_j(x_j) - \psi_k(x_k) + \kappa_{jk}}{\psi_i(x_i) - \psi_j(x_j) 
	+ \kappa_{ij}} = \frac{\chi_{jk}(x_j,x_k)}{\chi_{ij}(x_i,x_j)} 
	\: , \:\:\:\:\: k = 1, \ldots ,n \: ; \:\:\: k \neq i,j
\end{equation}
{\em Moreover, every Casimir invariant in (\ref{z1cas}) is globally defined in $\Omega$ and $C^{\infty}(\Omega)$.}

\mbox{} 

\noindent{\bf Proof.} Given that functions $\eta(\mathbf{x})$ and $\varphi_i(x_i)$ are nonvanishing in $\Omega$, the use of rank-preserving matrix transformations shows that Rank(${\cal J}$) $=$ Rank(${\cal X}$) in $\Omega$, where ${\cal X} \equiv (\chi_{ij}(x_i,x_j))$ for every pair $(i,j)$. Since at least one of the entries of ${\cal X}$ is also nonvanishing in $\Omega$, this implies that Rank(${\cal J}$) $\geq 2$ everywhere in $\Omega$. We can now submit matrix ${\cal X}$ to additional rank-preserving transformations: notice that Rank(${\cal X}$) is also maintained if we substract the first row to the rest of rows, and then if on the resulting matrix we substract the first column to every one of the remaining columns. This leads to a new matrix ${\cal X}^*$ given by:
\begin{equation}
\label{z1xbis}
	{\cal X}^* = \left( \begin{array}{cccc}
				0     &  \chi_{12}  &  \ldots  &  \chi_{1n}  \\
			- \chi_{12} &       0     &  \ldots  &     0       \\
			  \vdots    &   \vdots    &  \ddots  &  \vdots     \\
			- \chi_{1n} &       0     &  \ldots  &     0       \\
			 \end{array} \right)
\end{equation}
It is then clear from (\ref{z1xbis}) that Rank(${\cal J}$) $=$ Rank(${\cal X}^*$) $\leq 2$ in every point of $\Omega$. Therefore we conclude that Rank(${\cal J}$) $=2$ in $\Omega$. This demonstrates the first part of the statement. For the second part, notice first that every function $C_k(\mathbf{x})$ in (\ref{z1cas}) always depends on $x_i$, $x_j$ and $x_k$ (since functions $\psi_k(x_k)$ cannot be constant for any $k$, according to the conditions established) and in addition $C_k(\mathbf{x})$ does not depend on the rest of variables. This implies immediately the functional independence of the set $\{ C_k(\mathbf{x}): k= 1, \ldots ,n; \: k \neq i,j \}$. Moreover, since both $\chi_{jk}(x_j,x_k)$ and $\chi_{ij}(x_i,x_j)$ are $C^{\infty}(\Omega)$ and $\chi_{ij}(x_i,x_j) \neq 0$ everywhere in $\Omega$, function $C_k(\mathbf{x})$ is necessarily $C^{\infty}(\Omega)$. Therefore, to complete the proof it is only required to demonstrate that functions $C_k(\mathbf{x})$ are Casimir invariants for every $k$. The most simple way to see this is to verify that ${\cal J} \cdot \nabla C_k=0$ for every $k=1, \ldots , n$, with $k \neq i,j$ (notice that for both values $k=i,j$, function $C_k(\mathbf{x})$ is a constant, and then also a Casimir invariant, but trivial). We thus have: 
\[
	\partial _i C_k(\mathbf{x}) = \frac{\psi_i^{\prime} \chi_{kj}}{( \chi_{ij})^2} \: , \:\;\:
	\partial _j C_k(\mathbf{x}) = \frac{\psi_j^{\prime} \chi_{ik}}{( \chi_{ij})^2} \: , \:\;\:
	\partial _k C_k(\mathbf{x}) = \frac{\psi_k^{\prime} \chi_{ji}}{( \chi_{ij})^2} \: , \:\;\:
	k=1, \ldots n \: ; \:\:  k \neq i,j
\]
Then for every $r=1, \ldots ,n$ it can be seen that:
\begin{equation}
\label{z1t2jnc}
	\sum_{s=1}^n J_{rs} \partial _s C_k = J_{ri} \partial_i C_k + J_{rj} \partial_j C_k +
	J_{rk} \partial_k C_k = \frac{\eta \varphi_r}{(\chi_{ij})^2}
	(\chi_{ri}\chi_{kj} + \chi_{rj}\chi_{ik}+ \chi_{rk}\chi_{ji})
\end{equation}
In (\ref{z1t2jnc}) the last term vanishes for every choice of $i,j,k,r$, namely 
$\chi_{ri}\chi_{kj} + \chi_{rj}\chi_{ik}+ \chi_{rk}\chi_{ji} = 0$, as it was already shown in the proof of Theorem 1. Consequently, ${\cal J} \cdot \nabla C_k =0$ for every $k \neq i,j$. This completes the proof. \hfill Q.E.D.

\mbox{}

A consequence of the previous results is that they allow the constructive and global determination of the Darboux canonical form for this kind of Poisson structures: 

\mbox{}

\noindent {\bf Theorem 3.} 
{\em Let $\Omega \subset \mathbb{R}^n$ be a domain where a Poisson system (\ref{z1nham}) is defined everywhere, for which ${\cal J}$ is a structure matrix of the form (\ref{z1sol1}) characterized in Theorem 1, and such that the pair $(i,j)$ verifies that function $\chi_{ij}(x_i,x_j)$ is nonvanishing in $\Omega$. Then such Poisson system can be globally reduced in $\Omega$ to a one degree of freedom Hamiltonian system and the Darboux canonical form is accomplished globally in $\Omega$ in the new coordinate system $(y_1, \ldots ,y_n)$ and the new time $\tau$, where $(y_1, \ldots ,y_n)$ are given by the diffeomorphism globally defined in $\Omega$ 
\begin{equation}
\label{z1darbco}
	\left\{ \begin{array}{ccl}
	y_i (\mathbf{x}) & = & x_i  \\
	y_j (\mathbf{x}) & = & x_j  \\
	y_k (\mathbf{x}) & = & C_k(\mathbf{x}) \;\: , \;\:\;\: k=1, \ldots ,n ; \:\:\; k \neq i,j
	\end{array} \right.
\end{equation}
in which the $C_k(\mathbf{x})$ are the Casimir invariants (\ref{z1cas}); and the new time $\tau$ is defined by the time reparametrization:
}
\begin{equation}
\label{z1darbntt}
	\mbox{d} \tau = J_{ij}(\mathbf{x}(\mathbf{y})) \mbox{d} t
\end{equation}

\mbox{}

\noindent{\bf Proof.} Note that, according to Theorem 2, the Darboux theorem is applicable because ${\cal J}$ has constant rank 2 in $\Omega$. For the sake of clarity and without loss of generality, assume that it is $\chi_{12} \neq 0$ everywhere in $\Omega$. Recall also that, after a general smooth change of coordinates $\mathbf{y} = \mathbf{y}(\mathbf{x})$, an arbitrary structure matrix ${\cal J}(\mathbf{x})$ is transformed into another one ${\cal J^*}(\mathbf{y})$ as:
\begin{equation}
\label{z1jdiff}
      J^*_{ij}(\mathbf{y}) = \sum_{k,l=1}^n \frac{\partial y_i}{\partial x_k} J_{kl}(\mathbf{x}) 
	\frac{\partial y_j}{\partial x_l} \;\; , \;\:\; i,j = 1, \ldots , n
\end{equation}
For the family of interest, the reduction is carried out in two steps. We first perform the change of variables (\ref{z1darbco}), which in this case is
\begin{equation}
\label{z1d12}
	y_1 (\mathbf{x}) = x_1 \:\; , \;\:\; y_2 (\mathbf{x}) = x_2 \:\; , \;\:\; 
	y_k (\mathbf{x}) = C_k(\mathbf{x}) \;\: , \;\:\;\: k=3, \ldots ,n 
\end{equation}
where the $C_k(\mathbf{x})$ are given by (\ref{z1cas}), namely:
\begin{equation}
	C_k(\mathbf{x}) = \frac{\chi_{2k}(x_2,x_k)}{\chi_{12}(x_1,x_2)}
	= \frac{\psi_2(x_2)- \psi_k(x_k)+ \kappa_{2k}}{\psi_1(x_1) - \psi_2(x_2)+ \kappa_{12}}
	\;\: , \;\:\;\: k=3, \ldots ,n 
\end{equation}
Notice that this change of variables is invertible everywhere in $\Omega$, its inverse being:
\begin{equation}
\label{z1invd12}
	x_1 (\mathbf{y}) = y_1 \:\; , \;\:\; x_2 (\mathbf{y}) = y_2 \:\; , \;\:\; 
	x_k (\mathbf{y}) = \zeta_k(\psi_2(y_2)+\kappa_{2k}-y_k \chi_{12}(y_1,y_2)) 
		\;\: , \;\:\;\: k=3, \ldots ,n 
\end{equation}
where function $\zeta_k$ is the inverse function of $\psi_k$ for every $k$. The examination of (\ref{z1d12}-\ref{z1invd12}) shows that the variable transformation (\ref{z1d12}) to be performed exists and is a diffeomorphism everywhere in $\Omega$ as a consequence that by hypothesis we have $\chi_{12}(x_1,x_2) \neq 0$ and $\psi^{\prime}_k(x_k) \neq 0$ in $\Omega$. Then, according to (\ref{z1cas}) and (\ref{z1d12}), and taking (\ref{z1jdiff}) into account, after some algebra we are led to
\begin{equation}
\label{z1jdarb1}
	{\cal J^*}(\mathbf{y}) = J_{12}(\mathbf{x}(\mathbf{y})) \left( \begin{array}{cc}
		 0 & 1 \\ -1 & 0 \\ \end{array} \right) \oplus O_{(n-2) \times (n-2)}\end{equation}
where $O_{(n-2) \times (n-2)}$ denotes the null ${(n-2) \times (n-2)}$ matrix, and from (\ref{z1sol1}) and (\ref{z1invd12}) we have
\begin{equation}
\label{z1j12ntt}
	J_{12}(\mathbf{x}(\mathbf{y})) = \eta (y_1,y_2,x_3(\mathbf{y}), \ldots , x_n(\mathbf{y})) \varphi_1(y_1) \varphi_2(y_2) 
				\chi_{12}(y_1,y_2)
\end{equation}
The explicit dependences of $( x_3(\mathbf{y}), \ldots , x_n(\mathbf{y}) )$ are obviously the ones given in (\ref{z1invd12}) and were not displayed in (\ref{z1j12ntt}) for the sake of clarity. Note that $J_{12}(\mathbf{x}(\mathbf{y}))$ is nonvanishing in $\Omega ^* = \mathbf{y}(\Omega)$ and $C^{\infty}(\Omega ^*)$. These properties allow the accomplishment of the second step of the reduction which is a reparametrization of time, which in this case does not suppress the Poisson structure of the vector field. Thus, making use of (\ref{z1j12ntt}) in equation (\ref{z1darbntt}), the transformation $\mbox{d} \tau = J_{12}(\mathbf{x}(\mathbf{y})) \mbox{d} t$ is performed. This leads from the structure matrix (\ref{z1jdarb1}) to the Darboux canonical one:
\[
	{\cal J}_D (\mathbf{y}) = \left( \begin{array}{cc}
		 0 & 1 \\ -1 & 0 \\ \end{array} \right) \oplus O_{(n-2) \times (n-2)}
\]
Therefore the reduction is globally completed.  \hfill Q.E.D.

\mbox{}

\noindent {\bf Corollary 1.} 
{\em Every Poisson system defined in a domain $\Omega \subset \mathbb{R}^n$ in which the structure matrix is of the kind (\ref{z1sol1}) given in Theorem 1, is an algebraically integrable system in $\Omega$ and can be reduced globally and diffeomorphically in $\Omega$ to a Liouville integrable Hamiltonian system.
}

\mbox{}

The reader is referred to [\ref{gor1}] for the definitions of integrability used in Corollary 1. This concludes the analysis of the family of Poisson structures, since at this stage the reduction connects globally and constructively the initial Poisson systems with their classical Hamiltonian formulation. In what follows, the results developed are briefly illustrated by means of some applied examples. 

\mbox{}

\noindent{\bf Example 1.} Consider a domain $\Omega \subset \mathbb{R}^n$ in which the Poisson structure is to be defined, together with a generic function $\eta (\mathbf{x})$ which is smooth and nonvanishing in $\Omega$. Moreover, set $\varphi_i(x_i)=1$ and consistently $\psi_i(x_i)=x_i$ for every $i=1, \ldots ,n$, as well as $\kappa_{ij}=0$ for every pair $(i,j)$. This leads to: 
\begin{equation}
\label{z1jhalpn}
	J_{ij}(\mathbf{x}) = \eta (\mathbf{x})(x_i - x_j) \;\: , \;\:\;\: i,j=1, \ldots ,n
\end{equation}
Poisson structures of this form have deserved some attention for the study of the Halphen system [\ref{gyn1}] in which $\eta (x_1,x_2,x_3) = [2(x_1-x_2)(x_2-x_3)(x_3-x_1)]^{-1}$, 
as well as for the Poisson formulation of the system of circle maps [\ref{gyn1}] this time with
$\eta (x_1,x_2,x_3) = -[(x_1-x_2)(x_2-x_3)(x_3-x_1)]^{-1}$, both with $n=3$. Thus, 
(\ref{z1jhalpn}) is a natural $n$-dimensional generalization of the previous two Poisson structures. In order to fully comply with the requirements of Theorem 1 (and necessarily for the application of Theorems 2 and 3) it must be also assumed in (\ref{z1jhalpn}) that there exists at least one pair of indexes $(i,j)$ for which $\chi_{ij}(x_i,x_j)=x_i-x_j \neq 0$ everywhere in $\Omega$. Consistently with the previous style, this can be the case for 
$\chi_{12}$. Therefore, according to (\ref{z1cas}) and Theorem 2 a complete set of $C^{\infty}(\Omega)$ and functionally independent Casimir invariants associated to the structure matrices (\ref{z1jhalpn}) is:
\[
	C_k(\mathbf{x}) = \frac{x_2-x_k}{x_1-x_2} \;\: , \;\:\;\: k=3, \ldots ,n
\]
The rest of the Darboux reduction proceeds according to Theorem 3 without relevant difficulties. 

\mbox{}

\noindent{\bf Example 2.} Let $\Omega \subset \mathbb{R}^n$ be a domain such that $x_i \neq 0$ for every $\mathbf{x} \in \Omega$ and for every $i=1, \ldots ,n$. Also, consider a set of $n$ nonvanishing real constants $( \alpha_1 , \ldots , \alpha_n )$, their product being termed 
$\alpha = \displaystyle{\prod_{k=1}^n} \alpha_k \neq 0$, as well as a $C^{\infty}(\Omega)$ and nonvanishing function $\eta (\mathbf{x})$. We then define:
\begin{equation}
\label{z1jtopnd}
	J_{ij}(\mathbf{x}) = \eta (\mathbf{x}) ( \alpha_j x_i^2 - \alpha_i x_j^2) 
	\prod_{\scriptsize \begin{array}{c} k=1 \\ k \neq i,j \end{array}}^n x_k  
	\:\;\:\; , \:\;\:\;\:\;\:\; i,j=1, \ldots ,n
\end{equation}
If we assume for instance (and without loss of generality) that $( \alpha_2 x_1^2 - \alpha_1 x_2^2) \neq 0$ in $\Omega$, then it is not difficult to check that this is a structure matrix of the form characterized in Theorem 1. Under these assumptions, Poisson systems with structure matrix of the kind (\ref{z1jtopnd}) have the following set of functionally independent Casimir invariants in $\Omega$:
\[
	C_{k}(\mathbf{x}) = \frac{\alpha_1 \alpha_k x_2^2 - \alpha_1 \alpha_2 x_k^2}{\alpha_2 \alpha_k 
	x_1^2 - \alpha_1 \alpha_k x_2^2} \:\; , \:\;\:\;\:\; k=3, \ldots ,n
\]
The construction of the Darboux canonical form then follows the steps given in Theorem 3. For the sake of illustration from an applied perspective, note that the structure matrices 
(\ref{z1jtopnd}) generalize those arising in the case $n=3$ and $\eta(\mathbf{x})=1$, in which (\ref{z1jtopnd}) reduces to the following cubic structure matrix appearing [\ref{gyn1}] in the analysis of the Euler equations for a triaxial top:
\begin{equation}
\label{z1jtop}
	J_{ij}(x_1,x_2,x_3) = ( \alpha_j x_i^2 - \alpha_i x_j^2) \sum_{k=1}^3 
	(\epsilon_{ijk})^2 x_k \;\; , \;\;\;\; i,j = 1,2,3
\end{equation}
In (\ref{z1jtop}) $\epsilon_{ijk}$ is the Levi-Civita symbol, and $(\alpha_1,\alpha_2,\alpha_3)$ are real constants (related to the principal moments of inertia of the top) that are assumed to verify $\alpha _1 \alpha _2 \alpha _3 \neq 0$. 

\mbox{}

\begin{center}
{\bf REFERENCES}
\end{center}
\begin{enumerate}
   \item \label{olv1} P. J. Olver, {\em Applications of Lie Groups to Differential 
	Equations,\/} Springer-Verlag, New York (1993).
   \item \label{bsn1} B. Hern\'{a}ndez-Bermejo, New solution family of the Jacobi equations: 	characterization, invariants, and global Darboux analysis, {\em J. Math. Phys.} {\bf 	48} 022903 1-11 (2007). 
   \item \label{gyn1} H. G\"{u}mral and Y. Nutku, Poisson structure of dynamical systems with 	three degrees of freedom, {\em J. Math. Phys.} {\bf 34} 5691-5723 (1993).
   \item \label{byq} G. B. Byrnes, F. A. Haggar and G. R. W. Quispel, Sufficient conditions for 
	dynamical systems to have pre-symplectic or pre-implectic structures, {\em Physica\/}
	{\bf 272A} 99-129 (1999). 
   \item \label{byv4} B. Hern\'{a}ndez-Bermejo and V. Fair\'{e}n, Separation of variables in 	the Jacobi identities, {\em Phys. Lett.} {\bf 271A} 258-263 (2000). 
   \item \label{gor1} A. Goriely, {\em Integrability and Nonintegrability of Dynamical 
	Systems,\/} World Scientific, Singapore (2001).
\end{enumerate}
\end{document}